\documentclass[twocolumn,final]{aastex6}
\usepackage{float}
\usepackage{amsmath}
\usepackage{booktabs}
\usepackage{booktabs,threeparttable}
\usepackage{underscore}
\usepackage{lineno}
\usepackage{lipsum}
\turnoffedit

\begin{document}

\slugcomment{Submitted to \apj, November 2021}

\shorttitle{Observations of the Abell 98 Intercluster Filament}


\title{{\it Suzaku} Observations of the Cluster Outskirts and Intercluster Filament in the Triple Merger Cluster Abell~98}
\shortauthors{G. E. Alvarez, S. W. Randall, Y. Su et al.}

\author{Gabriella E. Alvarez}
\affiliation{Center for Astrophysics $|$ Harvard \& Smithsonian, 60 Garden Street, Cambridge, MA 02138}
\affiliation{Astrophysics Department, Vanderbilt University, Nashville, TN 37235}

\author{Scott W. Randall}
\affiliation{Center for Astrophysics $|$ Harvard \& Smithsonian, 60 Garden Street, Cambridge, MA 02138}

\author{Yuanyuan Su}
\affiliation{Physics and Astronomy, University of Kentucky, 505 Rose street, Lexington, KY 40506, USA}

\author{Arnab Sarkar}
\affiliation{Center for Astrophysics $|$ Harvard \& Smithsonian, 60 Garden Street, Cambridge, MA 02138}
\affiliation{Physics and Astronomy, University of Kentucky, 505 Rose street, Lexington, KY 40506, USA}

\author{Stephen Walker}
\affiliation{Department of Physics and Astronomy, The University of Alabama in Huntsville, Huntsville, AL 35899, USA}

\author{Nicholas P. Lee}
\affiliation{Center for Astrophysics $|$ Harvard \& Smithsonian, 60 Garden Street, Cambridge, MA 02138}

\author{Craig L. Sarazin}
\affiliation{Department of Astronomy, University of Virginia,
530 McCormick Road, Charlottesville, VA 22904-4325, USA}

\author{Elizabeth Blanton}
\affiliation{Institute for Astrophysical Research and the Department of Astronomy, Boston University, Boston, Massachusetts 02215, USA}

\begin{abstract}
 We present {\it Suzaku} observations of the Abell 98 (A98) triple galaxy cluster system and the purported intercluster filament. The three subclusters are expected to lie along a large scale cosmic filament. With partial azimuthal coverage of the northernmost cluster, we find that the inferred entropy profile of this relatively low mass cluster ($kT \approx 2.8$ keV) adheres to expectations from models of self-similar pure gravitational collapse in the region of the virial radius. There is evidence of extended structure beyond $r_{200}$ to the north of the northernmost cluster, along the merger axis, with properties consistent with what is expected for the warm-hot intergalacitc medium (WHIM; $kT = 0.11_{-0.02}^{+0.01}$ keV, $n_e = {7.6 \times 10^{-5}}^{+3.6 \times 10^{-5}}_{-3.6 \times 10^{-5}}$~cm$^{-3}$). No such emission is detected at the same radius in regions away from the merger axis, consistent with the expectation that the merger axis of this triple system lies along a large scale cosmic filament. In the bridge region between A98N and A98S, there is evidence of filamentary emission at the $2.2\sigma$ level, as well as a tentative detection of cool gas ($kT \sim 1$ keV). The entropy profile of this intercluster filament suggests that the A98 system is most likely aligned closer to the plane of the sky rather than along the line of sight. The structure to the north of the system, as well as in between A98N and A98S is indicative that the clusters are connected to a larger-scale structure spanning at least 4 Mpc.
\end{abstract}


\maketitle

\section{Introduction} \label{sec:intro}

Cosmological simulations predict that massive galaxy clusters are generally located at the nodes of the web-like, large-scale structure of the universe (e.g. IllustrisTNG simulation; \cite{nelson}). The assembly and evolution of galaxy clusters throughout time are powerful tools for precision cosmology (e.g. \cite{schellenberger}), which informs our understanding of the growth and evolution of structure. Simulations predict that the baryons in the cosmic web have been shock heated and compressed to electron temperatures $T_e$~$\approx 10^{5} - 10^{7}$~K with electron densities $n_{e} \approx 10^{-7} - 10^{-5}$~cm$^{-3}$. This diffuse, primordial baryonic gas is known as the warm-hot intergalactic medium (WHIM), which is thought to comprise \deleted{a significant portion} \edit1{approximately $50\%$} of baryons in the local universe \citep[e.g.,][]{bregman}. 

One key assumption when using galaxy clusters for precision cosmology is that clusters are in hydrostatic equilibrium. However, it has been observed that many massive galaxy clusters seemingly deviate from hydrostatic equilibrium at large cluster radii (e.g. \cite{walker2013}). This could be due to a number of physical processes in the intracluster medium (ICM), including clumping of cool gas \citep[][]{eckert2015,simionescu,tchernin} at large cluster radii ($\sim r_{200}$  \footnote{\deleted{$r_\Delta$ is the radius at with the density of the intracluster medium is $\Delta$ times the critical density of the universe.} \edit1{$r_\Delta$is the radius of a sphere whose mean matter density is $\Delta$ times the critical density of the universe}}) biasing the average surface brightness of cluster outskirts high, non-thermal pressure support from bulk motions, turbulence or cosmic rays \citep[][]{lau,vazza,battaglia}, and \deleted{electrion}\edit1{electron}-ion non-equilibrium \citep[][]{fox,wong2009,hoshino,avestruz} (for a review of cluster outskirts, see e.g. \cite{walker2019}).

The ICM thermodynamic history of clusters is encoded in the entropy profiles of the ICM. Unlike larger mass clusters, studies of lower mass galaxy clusters and groups with {\it Suzaku} have shown close agreement in the virialization region between the predicted entropy profiles based on self-similar gravitational collapse models, and the inferred entropy profiles of the systems based on X-ray spectroscopy studies \citep[][]{su,bulbul,sarkar}. This could be due to the fact that galaxy groups are likely more evolved than galaxy clusters \citep[][]{paul}. 

Another theory is that due to their smaller mass, galaxy groups are more sensitive to common thermodynamic processes such as stellar and active galactic nuclei (AGN) feedback \citep[e.g.][]{pratt2010,lovisari}, injecting entropy into their outskirts. This injected entropy could artificially correct for the entropy deficiency observed in more massive clusters around the virialization region, making the ICM of the lower mass cluster or group look like it's obeying self-similarity. However, an entropy excess at large group and low mass cluster radii has not yet been observed. Isolated groups also tend to be in less dynamically active regions than clusters, as they are not at the nodes of the cosmic web. The relative isolation would lead there to be fewer clumps, weaker accretion shocks, and less turbulence.

Simulations \citep[e.g.][]{angelinelli} and deep X-ray observations \citep[e.g.][]{mirakhor} of galaxy clusters have revealed that the thermodynamic structure of the outskirts varies azimuthally, presumably due in large part to the growth of hierarchical structure through cosmic filaments. 
Understanding the physical processes at work in the ICM in the outskirts of clusters is key to our \deleted{understating} \edit1{understanding} of the growth of cosmic structure, and to using clusters for precision cosmology.

\begin{table}[ht!]
\caption{J2000 right ascension and declination, redshift ($z$), electron temperature ($T_e$), and $r_{500}$ for the three main subclusters consisting of the A98 system.}
\label{tab:objects}
\begin{tabular}{@{}cccccc@{}}
\toprule
Object & R.A.                                & Dec                                 & z      & $T_e$~[keV] & $r_{500}$~[Mpc] \\ \midrule
A98N   & $00^{\rm{h}}46^{\rm{m}}25^{\rm{s}}$ & $20^{\rm{d}}37^{\rm{m}}14^{\rm{s}}$ & 0.1043 & 2.78                  & 0.72                      \\
A98S   & $00^{\rm{h}}46^{\rm{m}}29^{\rm{s}}$ & $20^{\rm{d}}28^{\rm{m}}04^{\rm{s}}$ & 0.1063 & 2.82                  & 0.72                      \\
A98SS  & $00^{\rm{h}}46^{\rm{m}}36^{\rm{s}}$ & $20^{\rm{d}}15^{\rm{m}}44^{\rm{s}}$ & 0.1218 & 2.42                  & 0.67                      \\ \bottomrule
\end{tabular}
\end{table}

Abell 98 consists of three subclusters \citep[][]{abell}; Abell 98N (A98N), Abell 98S (A98S), and Abell 98SS (A98SS). The right ascension and  declination \citep[][]{jones1999,burns}, redshift \citep[][]{white,pinkney}, and measured electron temperature and inferred $r_{500}$ \citep[][]{vikhlinin} are presented in Table~\ref{tab:objects}.

The bimodal distribution of A98N and A98S was first seen in the X-ray with the {\it Einstein} telescope \citep[][]{forman,henry}. A98SS was also later observed with the {\it Einstein} telescope \citep[][]{jones1999}. The presence of three subclusters aligned colinearly on Mpc scales suggests the presence of a local large scale structure filament aligned with the subclusters.

\cite{paterno-mahler} found, with relatively shallow {\it Chandra} and XMM-{\it Newton} observations, that there is evidence for a shock heated region to the south of A98N, perhaps due to an early stage cluster merger.  Their dynamical analysis supported this conclusion. This study also determined that A98S has an asymmetric ICM distribution, indicating that A98S is still undergoing a major merger.

\cite{sarkarA98} found definitive evidence of a shock edge to the south of A98N with deeper {\it Chandra} observations, confirming that A98N and A98S are likely in the early stages of a merging event.

The errors reported are $90\%$ unless otherwise stated. Throughout this work, we assume $H_0 = 70$~km s$^{-1}$ Mpc$^{-1}$, $\Omega_M$=0.3, and $\Omega_\Lambda$=0.7. The average redshift of the A98 system is $\bar{z}=0.1042$ corresponding to 1$\arcsec=1.913$~kpc. For this analysis we assume the abundance table of \cite{grevesse}.

\begin{table*}[t]
\centering
\caption{Details of the {\it Suzaku}, {\it Chandra}, and XMM-{\it Newton} observations used in this work.}
\label{tab:observations}
\begin{tabular}{@{}cccccccc@{}}
\toprule
Observatory     & Pointing   & ObsID     & RA                                       & Dec                                      & Date Obs   & \begin{tabular}[c]{@{}c@{}}Exposure {[}ks{]}\\ ACIS-I\\ XIS0/XIS1/XIS3 \\ MOS1/MOS2/PN \end{tabular} & PI         \\ \midrule
{\it Suzaku}  & A98C & 809077010 & 00$^{\rm h}$46$^{\rm m}$ 29$^{\rm s}$.93 & +20$^{\rm h}$33$^{\rm m}$ 25$^{\rm s}$.9 & 2014-06-09 & 59.75/58.86/59.63                                                                   & S. Randall \\
{\it Suzaku}  & A98N       & 809078010 & 00$^{\rm h}$46$^{\rm m}$ 18$^{\rm s}$.12 & +20$^{\rm h}$50$^{\rm m}$ 20$^{\rm s}$.0 & 2014-07-02 & -/-/24.9                                                                            & S. Randall \\
{\it Suzaku}  & A98N       & 809078020 & 00$^{\rm h}$46$^{\rm m}$ 18$^{\rm s}$.02 & +20$^{\rm h}$50$^{\rm m}$ 21$^{\rm s}$.1 & 2014-07-03 & 23.3/21.1/22.5                                                                      & S. Randall \\
{\it Suzaku}  & A98N       & 809078030 & 00$^{\rm h}$46$^{\rm m}$ 19$^{\rm s}$.18 & +20$^{\rm h}$49$^{\rm m}$ 43$^{\rm s}$.3 & 2014-12-21 & 48.9/50.3/51.5                                                                      & S. Randall \\
{\it Suzaku}  & A98S       & 809080010 & 00$^{\rm h}$46$^{\rm m}$ 42$^{\rm s}$.48 & +20$^{\rm h}$16$^{\rm m}$ 30$^{\rm s}$.0 & 2014-12-20 & 47.7/48.4/49.0                                                                      & S. Randall \\
{\it Suzaku}  & A98W       & 809079010 & 00$^{\rm h}$45$^{\rm m}$ 15$^{\rm s}$.40 & +20$^{\rm h}$40$^{\rm m}$ 10$^{\rm s}$.2 & 2014-07-03 & 95.6/96.9/99.7                                                                      & S. Randall \\
{\it Chandra} & A98N       & 11877     & 00$^{\rm h}$46$^{\rm m}$ 24$^{\rm s}$.80 & +20$^{\rm h}$28$^{\rm m}$ 05$^{\rm s}$.0 & 2009-09-17 & 19.5                                                                                & S. Murray     \\
{\it Chandra} & A98S       & 11876     & 00$^{\rm h}$46$^{\rm m}$ 29$^{\rm s}$.30 & +20$^{\rm h}$37$^{\rm m}$ 17$^{\rm s}$.0 & 2009-09-17 & 19.8                                                                                & S. Murray     \\
{\it Chandra} & A98SS      & 12185     & 00$^{\rm h}$46$^{\rm m}$ 36$^{\rm s}$.10 & +20$^{\rm h}$15$^{\rm m}$ 22$^{\rm s}$.5 & 2010-09-08 & 19.8                                                                                & S. Murray     \\ 
XMM-{\it Newton} & A98C & 0652460201 & 00$^{\rm h}$46$^{\rm m}$ 13$^{\rm s}$.40 & +20$^{\rm h}$34$^{\rm m}$ 47$^{\rm s}$.5 & 2010-12-26 & 33.0/33.0/24.0 & C. Jones \\ \bottomrule

\end{tabular}
\end{table*}

\section{Observations and Data Reduction} \label{sec:datareduction}

In this Section we discuss the data and reduction techniques for the X-ray observations of A98 listed in Table~\ref{tab:observations}.

\subsection{Suzaku} \label{subsec:datareductionSuzaku}

{\it Suzaku} observed the A98 system with a total of 6 pointings (see Table~\ref{tab:observations}) from June-December 2014. Three of the four XIS detectors, XIS0, XIS1, and XIS3, were on for each observation. 
We find that  these {\it Suzaku} observations are offset by $\sim 33$\arcsec~in right ascension from both the {\it Chandra} and XMM-{\it Newton} observations of the system. When analyzing data from the different observations jointly, this astrometry correction is taken into account.

\begin{figure*}[ht!]
    \centering
    \includegraphics[width=\textwidth]{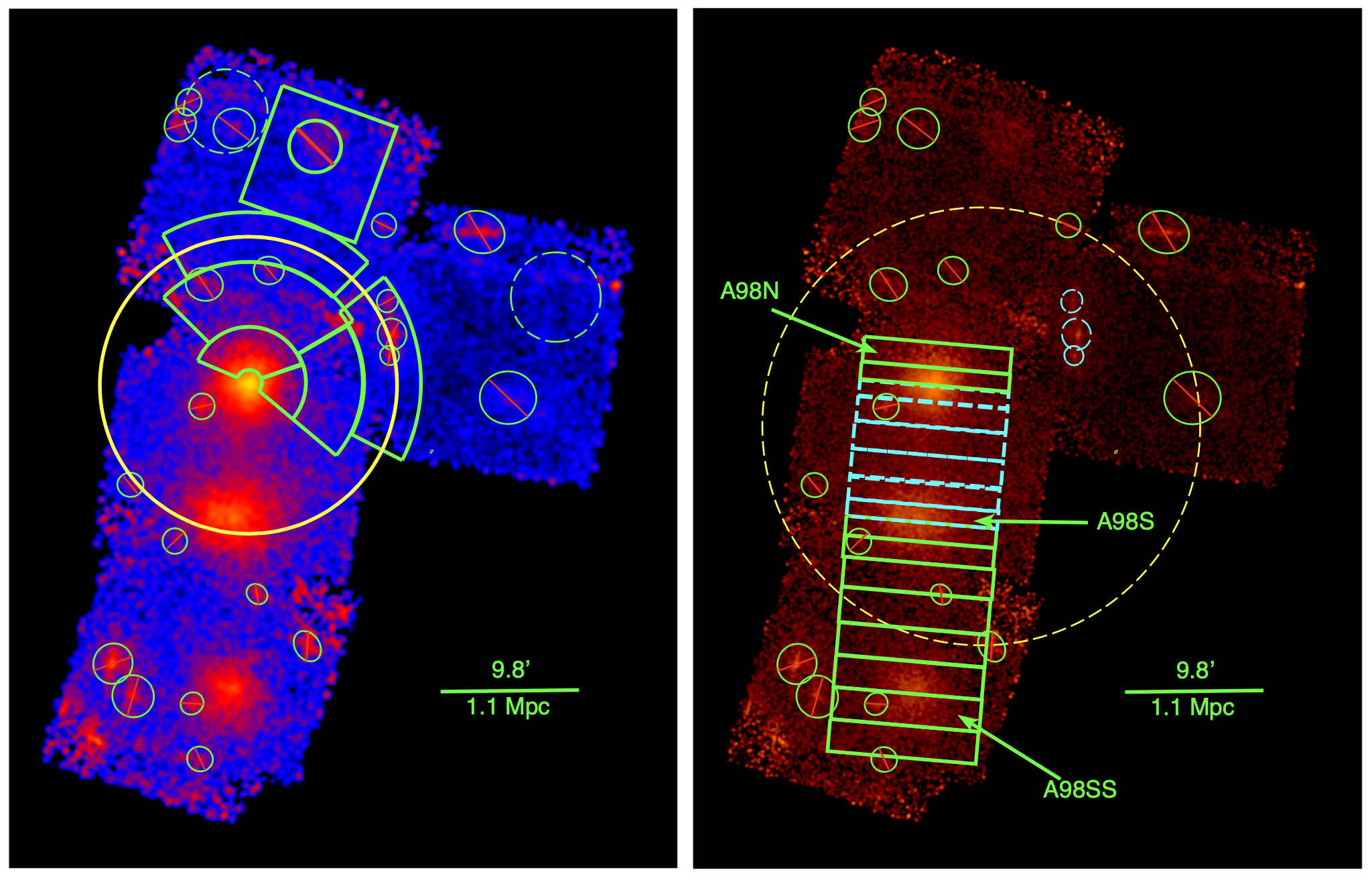}
    \caption{Left: Annotated, mosaic, background subtracted, and exposure corrected {\it Suzaku} image of the A98 system in the $0.5-7.0$~keV band. The image is binned to emphasize diffuse features in the system, and smoothed with a 20.8$\arcsec$~Gaussian. The sectors are the regions used for the temperature and entropy profiles for A98N. The yellow circle denotes $r_{200}$ for A98N derived from the $T_{500}$-$r_{500}$ relation found by \cite{vikhlinin}. The dashed green circles are regions used to constrain the sky background. The excluded regions are point sources, with the exception of a potential background cluster (see Section~\ref{subsec:filres}), and are excluded from further analysis.  The cyan dashed circle regions mark the point sources that are modelled simultaneously with the outermost sector to the west of A98N. The box region to the north is used to estimate large-scale properties described in Section~\ref{subsec:filres}. Right: The {\it Suzaku} image is smoothed with a 10.4$\arcsec$~circular Gaussian. The box regions are used for the intercluster filament thermodynamic profiles presented in Section~\ref{subsec:bridgeProp}. The dashed cyan boxes are used to derive the temperature, electron density, and entropy profile for A98N and A98S presented in Figure~\ref{fig:a98bridgeent}. The dashed yellow circle denotes the XMM-{\it Newton} field of view. The cyan dashed regions and exclusion regions are the same as in the left panel.}
    \label{fig:mosaic}
\end{figure*}

The {\it Suzaku} data are reduced using HEASOFT version 6.22.1 and the latest calibration database as of May 2014. \edit1{When CIAO tools are being used for analysis, CIAO version 4.9 and CALDB version 4.7.7 were employed}. The FTOOL {\em aepipeline} is used for the first order reprocessing of the unfiltered files.  Appropriate filters were applied for an Earth elevation greater than 5$^\circ$, a sunlit Earth elevation greater than 20$^\circ$, a cutoff rigidity greater than 6 GeV/c, and passages just before and after the South Atlantic Anomaly with {\em XSELECT}. The 5x5 editing mode files are first converted to the 3x3 editing mode using {\em xis5x5to3x3} and then the events are combined for each detector. The corners of the chips illuminated by the Fe$^{55}$ calibration sources are then removed. The second rows adjacent to the charge injected rows at 6 keV are removed from XIS1 for further analysis due to the increase in NXB level on XIS1. Due to the micrometeorite hit to XIS0, the impacted region is excluded from the XIS0 field throughout the analysis. Light curves are extracted from the events files with the CIAO tool {\em dmextract}. The CIAO tool {\em deflare} is used on the light curves with the iterative sigma clipping routine {\em lc_sigma_clip} set to $3\sigma$.  No instances of strong flaring in the observations are found.

The non-X-ray background (NXB) images are generated using the {\em xisnxbgen} routine \citep{tawa}. The NXB images are then scaled so that the hard-band (10-12 keV) count rate matches the source {\it Suzaku} observations. After creating source and scaled NXB mosaic images, the mosaic NXB image is then subtracted  from the source image. Flat field images are generated with the routine {\em xissim} to create effective exposure maps. \deleted{The mosaic exposure map is divided by the NXB subtracted source mosaic image} \edit1{The mosaic NXB image is subtracted from the mosaic source image, and the resulting image is divided by the mosaic exposure map} to create the final image shown in Figure~\ref{fig:mosaic}. \edit1{The PSF of {\it Suzaku} is relatively large ($\sim 2\arcmin$)}. The CIAO tool {\em wavdetect} is used on wavelet scales of 14, 28, and 56 pixels, where each pixel is $2\arcsec$ in length after binning, to detect point sources. \edit1{These scales are roughly 1/4, 1/2, and 1 times the size of the $2\arcmin$ PSF, which should detect both bright point sources and fainter point sources where the broad PSF wings may be too faint to be detectable.} The image was then inspected by eye in order to make appropriate adjustments to the point source regions detected for exclusion from further analysis.

\subsection{Chandra} \label{subsec:dataReductionChandra}

The A98 system was observed with {\it Chandra}, with a total of three pointings from September 2009-September 2010 (see Table~\ref{tab:observations}). More recent, deeper {\it Chandra} obervations of this system will be presented in Sarkar et al. (in prep). CIAO version 4.9 and CALDB version 4.7.7 are used for the {\it Chandra} data reduction. After using {\em chandra_repro} to generate level 2 events files (see Table~\ref{tab:observations} for observations), light curves are extracted with the CIAO tool {\em dmextract}, and then run through the {\em deflare} routine to filter the level 2 events files for any possible flares during the observation.

\begin{figure}[ht!]
    \centering
    \includegraphics[width=0.45\textwidth]{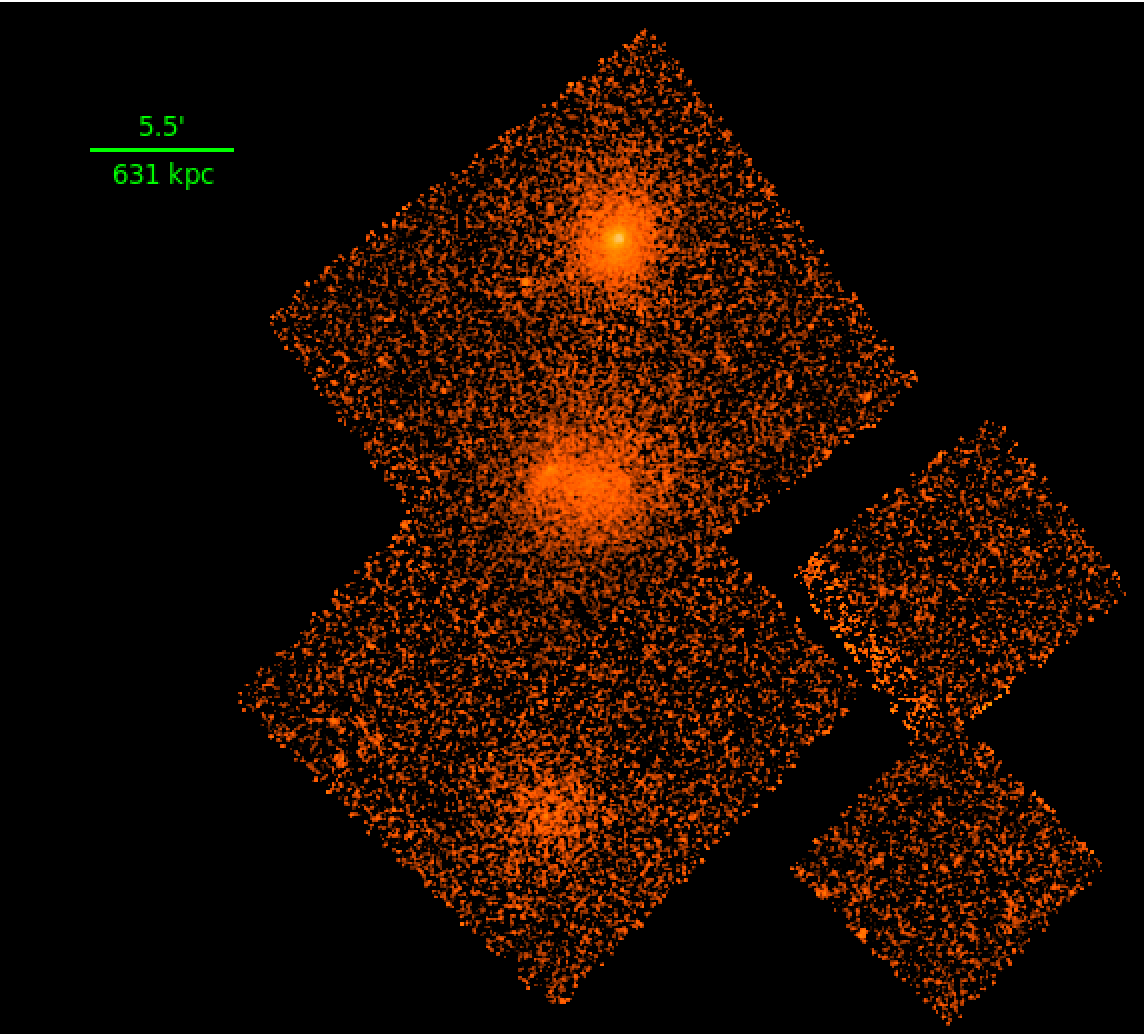}
    \caption{0.3-7.0 keV blank sky background subtracted, exposure corrected, mosaic image of the {\it Chandra} observations (see Table~\ref{tab:observations}). The image is smoothed with a 10\arcsec~ Gaussian.}
    \label{fig:chandraImg}
\end{figure}

The {\it asphist, mkinstmap,} and {\it mkexpmap} routines in CIAO are used to generate exposure maps for imaging. The blank sky background files with the closest observation periods to each of the {\it Chandra} observations are reprojected to match the observations using the CIAO routine {\em reproject\_events}. The blank sky background images are hard-band (10-12 keV) scaled to match high-energy particle background count rate of the source observations. We use the {\em wavdetect} routine on wavelet scales of 2, 4, 6, 8, 10, 12, and 16 pixels, where the pixels are 0.98\arcsec\ in length to detect background sources for removal. These sources are excluded from further analysis. The routine {\em dmfilth} is used to fill in the excluded source regions by drawing counts from a Poisson distribution matched to the local surrounding regions for imaging. The mosaic background subtracted, exposure corrected {\it Chandra} 0.3-7.0 keV image is presented in Figure~\ref{fig:chandraImg}.

This {\it Chandra} data is used to corroborate {\it Suzaku} measurements, and is in general agreement with the deeper observations presented in \cite{sarkarA98}.

\subsection{XMM-{\it Newton}} \label{subsec:dataReductionXMM}

The Source Analysis Software (SAS) v18.0.0 is used to reduce the XMM-{\it Newton} event lists for A98N. The tools {\em odfingest} and {\em cifbuild} were used to create \edit1{the Observation Data Files (ODF)} summary file and to build a \edit1{to build a CCF (Current Calibration File) Index File (CIF)} \deleted{file} respectively to prepare the data for further analysis. We then use the tool {\em xmmextractor} to generate images and calibrated events files for point-like source analysis. A smoothed 0.5-2.0 keV MOS2 image of this observation is presented in Figure~\ref{fig:xmmNewton}. 

\section{Analysis} \label{sec:analysis}

{\em XSPEC} version 12.11.0 was used to perform the spectral analysis. An absorbed (phabs) APEC model \citep{smith} was used for the source spectra.

\subsection{Suzaku Spectroscopy} \label{subsec:suzakuSpec}

To extract spectra for the {\it Suzaku} observations, the {\em XSELECT} environment and subsequently the {\em extract spec} routine are utilized. Then the corresponding redistribution matrix file (RMF) is generated with {\em xisrmfgen} . The RMF and source spectrum are then used to generate the ancillary response files (ARFs) with {\em xissimarfgen}.

Two different ARFs are generated for each spectrum to be folded into two different model components in the spectral fits; the source model and the background model. The {\it Chandra} image is used to fit a 2D-$\beta$ model for all three clusters. In order to create more accurate ARFs for the source spectra, {\em xissimarfgen} utilizes an image of the 2D-$\beta$ model. A uniform circle with a radius of $20\arcmin$ is used to generate the ARFs folded into the background model, under the assumption that it is uniform.

The background model includes components from the local hot bubble (LHB), the galactic halo (GH) and the cosmic X-ray background (CXB). Non X-ray background (NXB) spectra are generated with {\em xisnxbgen} (for more details see \citep{tawa}) and are subtracted from the source spectra in {\em XSPEC}. We group all of our spectra such that each bin contains a minimum of 40 counts. In order to further constrain the GH and LHB, and CXB emission, we extract a spectrum of an annulus from the {\it ROSAT} All Sky Survey ({\it RASS})\footnote{https://heasarc.gsfc.nasa.gov/cgi-bin/Tools/xraybg/xraybg.pl}. The {\it RASS} annulus is centered on RA = $0^{\rm h}46^{\rm m}18^{\rm s}.8872$ and Dec = $20^{\rm h}28^{\rm m}13^{\rm s}.557$, approximately on the X-ray centroid of A98SS, with an inner radius of $0.5643^\circ$ and an outer radius of $0.6738^\circ$. The free model parameters can be seen in Table~\ref{tab:bgparams}. The LHB temperature is fixed to $kT=0.1$~keV, \edit1{the GH temperature is fixed to $kT = 0.2$~keV} and the powerlaw component for the CXB is fixed to $\Gamma = 1.4$. \edit1{The normalization for the GH, LHB, and CXB are left as free parameters in all spectral fits presented in this work in order to account for systematic errors.}

\begin{table*}[t]
\centering
\caption{The free parameters for the fits to the dashed green background regions shown in Figure~\ref{fig:mosaic} (Left).}
\label{tab:bgparams}
\begin{tabular}{@{}ccccc@{}}
\toprule
BG Region & \begin{tabular}[c]{@{}c@{}}CXB Norm \\ {[}photons/keV/cm$^2$/s/arcmin$^2${]}\end{tabular} & GH $kT$ {[}keV{]}         & \begin{tabular}[c]{@{}c@{}}GH Norm \\ {[}cm$^{-5}$arcmin$^{-2}${]}\end{tabular} & \begin{tabular}[c]{@{}c@{}}LHB Norm \\ {[}cm$^{-5}$arcmin$^{-2}${]}\end{tabular} \\ \midrule
West      & $6.9e-7^{+4\rm{e}-8}_{-4\rm{e}-8}$                                                       & $0.19_{-0.01}^{+0.03}$ & $1.3e-6_{-3\rm{e}-7}^{+8\rm{e}-7}$                                                        & $2.1e-7_{-2\rm{e}-7}^{+1\rm{e}-7}$                                                                \\
North     & $6.8e-7^{+3\rm{e}-8}_{-3\rm{e}-8}$                                                       & $0.19_{-0.02}^{+0.04}$ & $1.4e-6_{-4\rm{e}-7}^{+8\rm{e}-7}$                                                        & $1.8e-7_{-1\rm{e}-7}^{+1\rm{e}-7}$                                                                \\ \bottomrule
\end{tabular}
\end{table*}

Most point sources are removed by inspecting the {\it Suzaku}, {\it Chandra}, and XMM-{\it Newton} observations listed in Table~\ref{tab:observations}. The bright point sources that are not excluded from the regions for analysis are modelled simultaneously. We first model the point sources with an absorbed powerlaw model for AGN, and an APEC model for a foreground galactic star, using the XMM-{\it Newton} data that overlaps the {\it Suzaku} field of view (FOV). We then take the parameters from the model to simulate the same {\it Suzaku} point source with the FTOOL {\it xissim}. We use this simulation to estimate the {\it Suzaku} normalization for the point source models and then fold the point source models into the region fits, allowing the point source parameters to vary within their $90\%$ errors derived from the XMM-{\it Newton} fit of the source.

\subsubsection{{\it Suzaku} Scattered Light}
\label{suzakuScatteredLight}

Due to {\it Suzaku's} relatively large PSF, one must consider the effects of scattered light from adjacent regions when performing spectral analysis.
In order to determine the effect of scattered light across adjacent regions, we use {\it xissim} to simulate each region with $2\times 10^6$ photons as is done in e.g.  \cite{walker2012} and \cite{bulbul}. We calculate the percentage of photons that originate from the region of interest and are detected in that region, and what percentage of photons are scattered into the adjacent region in order to include a properly weighted component in the spectral fit to the adjacent region (Table~\ref{tab:scatteredlight}). We use these values to test whether the scattered light affects our results, particularly in the faintest regions analyzed in this work. We find that the scattered light does not change the final results for the analysis of A98N, as the statistical error dominates the effect of scattered light. 
We find that the scattered light from outside of the FOV is negligible.  Therefore, we omitted values from Table~\ref{tab:scatteredlight} where the adjacent sector was outside of the FOV of the observation containing the source sector.

\begin{table}[]
\label{tab:scatteredlight}
\centering
\caption{Percentage of light scattered from the regions in the columns into the regions in the rows. ``N" represents the sectors to the north of A98N, and ``W" represents the sectors to the west of A98N.}
\begin{tabular}{@{}ccccccc@{}}
\toprule
Region & N01  & N02  & N03  & W01  & W02  & W03  \\ \midrule
N01    & 65.1 & -    & -    & -    & -    & -    \\
N02    & -    & 73.2 & 6.69 & -    & -    & -    \\
N03    & -    & 9.38 & 70.6 & -    & -    & -    \\
W01    & -    & -    & -    & 54.5 & 6.2  & -    \\
W02    & -    & -    & -    & 9.86 & 60.0 & -    \\
W03    & -    & -    & -    & -    & -    & 60.7 \\ \bottomrule
\end{tabular}
\end{table}

\subsubsection{Systematic Error in the Cosmic X-Ray Background} \label{subsec:susys}

{\it Suzaku} is able to detect point sources down to a flux of $1\times10^{-13}$ ergs cm$^{-2}$~s$^{-1}$~deg$^{-2}$. \cite{moretti} defines the unresolved CXB flux in ergs cm$^{-2}$~s$^{-1}$~deg$^{-2}$ as 
\begin{equation}
F_{CXB} = (2.18 \pm 0.13) \times 10^{-11} - \int_{S_{lim}}^{S_{max}} \frac{dN}{dS}\times S~d\deleted{s}\edit1{S}.
\end{equation}
\edit1{The normalization in this equation is derived from {\it SWIFT} data \citep{moretti}.} The analytical form of the source flux distribution in the 2-10 keV band is characterized as \citep{moretti}

\begin{equation}
N(>S) = N_0 \frac{(2 \times 10^{-15})^\alpha}{S^\alpha + S_0^{\alpha - \beta}S^\beta} ~ \rm{ergs}~\rm{cm}^{-2}~\rm{s}^{-1}, 
\end{equation}
where $\alpha = 1.57_{\deleted{-0.18}\edit1{-0.08}}^{+0.10}$ and $\beta = 0.44^{+0.12}_{-0.13}$ are the power laws for the bright and faint components of the distribution respectively, $N_0 = 5300_{-1400}^{+2850}$, \edit1{$S_0 = 4.5_{-1.7}^{+3.7} \times 10^{-15}$~ergs cm$^{-2}$ s$^{-1}$,} $S_{lim}$ is the flux of the faintest point source detected in the observation, and $S_{max} = 8 \times 10^{-12}$ ergs cm$^{-2}$~s$^{-1}$. The nominal value, $S_{lim}$, for {\it Suzaku} point sources is $1\times10^{-13}$ ergs cm$^{-2}$~s$^{-1}$~deg$^{-2}$ \citep[e.g.][]{walker2013}. Therefore the unresolved CXB in the background region used in the {\it Suzaku} observations has a flux of $1.87 \pm$ \deleted{0.13} $\edit1{0.23} \times 10^{-11}$ ergs cm$^{-2}$~s$^{-1}$~deg$^{-2}$. 

Finally, the expected $1\sigma$ \edit1{spatial variance} \deleted{uncertainty} in the unresolved CXB flux may be given by:

\begin{equation}
\sigma^2 = \frac{1}{\Omega}\int_{0}^{S_{lim}}\frac{dN}{dS}\times S^2~d\deleted{s}\edit1{S}
\end{equation}
where $\Omega$ is the solid angle \citep{bautz}. We use this method to constrain the normalization of the CXB component of the background, and allow this parameter to vary within the derived $1\sigma$ errors.

\subsection{XMM-Newton Spectral Analysis} \label{subsec:xmmSpec}

\begin{figure}[ht!]
    \centering
    \includegraphics[width=0.45\textwidth]{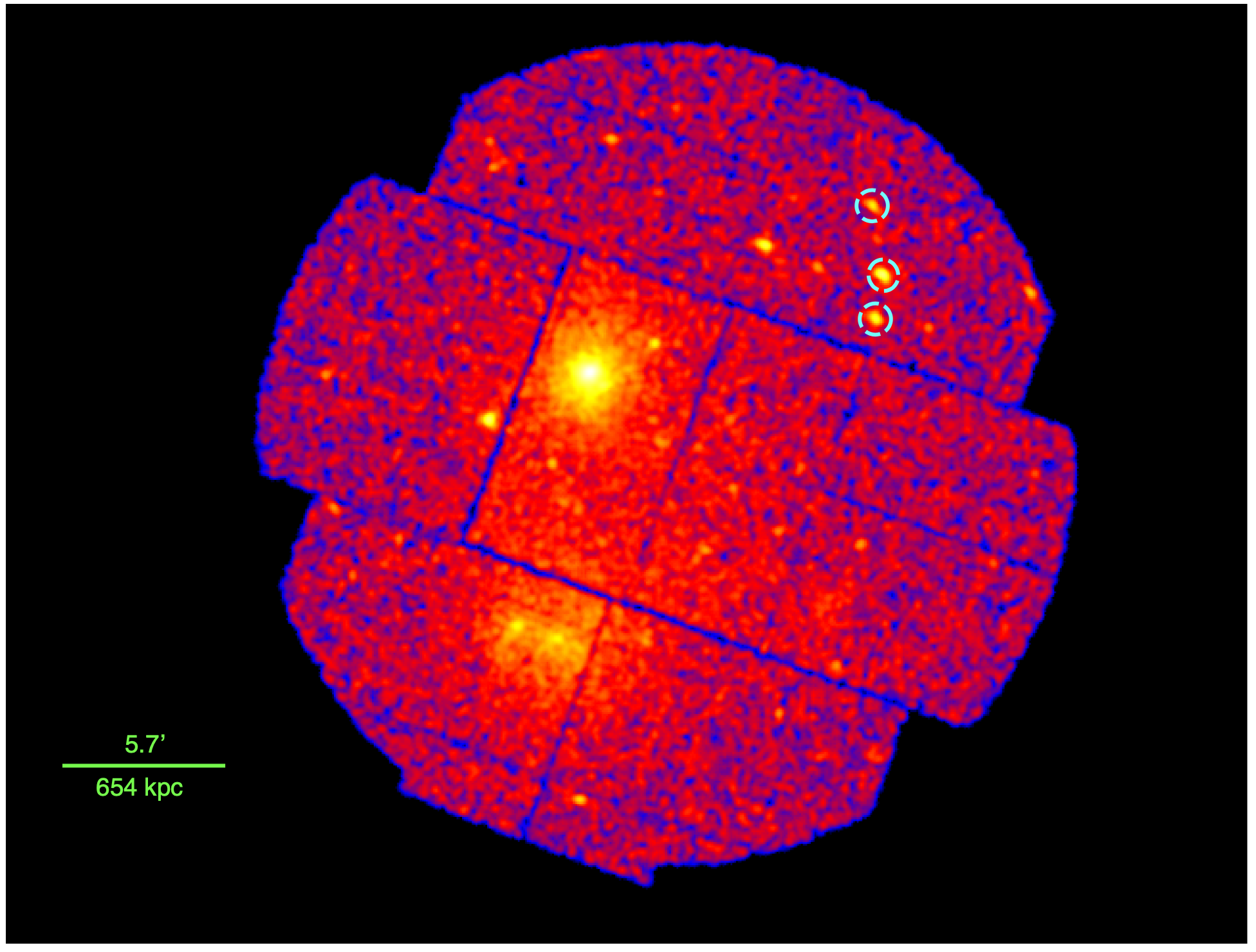}
    \caption{XMM-{\textit Newton} MOS2 image of A98N and A98S smoothed with a $15 \arcsec$ Gaussian kernel. The point sources modelled with XMM-{\textit Newton} data in this work are highlighted with dashed cyan regions. }
    \label{fig:xmmNewton}
\end{figure}

SAS is used to generate all of the necessary files of the point-like regions shown in Figure~\ref{fig:xmmNewton} for analysis in {\em XSPEC}. In order to generate the source and background spectra, we use the {\em evselect} routine. We then use the {\em rmfgen} routine to generate the redistribution matrix files and the {\em arfgen} routine to generate the ancillary response file for the spectrum. The XMM-{\it Newton} spectra are grouped such that each bin contains a minimum of 25 counts due to the relatively low counts observed for the point sources. We then use an absorbed powerlaw model in {\em XSPEC} for the point sources that are AGN, and a simple APEC model for the point sources that are galactic stars.

\section{Results and Discussion} \label{sec:resultsDiscuss}

In this section we present temperature, electron density, and entropy profiles for regions shown in Figure~\ref{fig:mosaic}.

The electron density of regions of interest is derived from the normalization for the APEC model in {\em XSPEC}. The electron density may be derived from the {\em XSPEC} normalization with the following equation:
\begin{equation}
    n_e  = \biggl[1.2 \ N \times 4.07\times10^{-10}(1+z)^2 \\ \biggl (\frac{D_A}{\rm{Mpc}}\biggl)^2\biggl(\frac{V}{\rm{Mpc^3}}\biggl)^{-1}\biggl]^{1/2},
\label{eq:density}
\end{equation}

\noindent where $D_A$ is the angular diameter distance to the system, $V$ is the volume for an assumed geometry of the region, $N$ is the {\em XSPEC} normalization, and $z$ is the redshift of the object. \edit1{The assumed ratio for $\frac{n_e}{n_H}$~is $1.2$ \citep{boehringer}.}

\subsection{Cluster Thermodynamic Properties} \label{subsec:tempRes}

Here we present thermodynamic profiles of the subclusters in the A98 system. In order to deproject our temperature measurements, we use the ``onion peeling" method \citep{ettori} and assume a spherical geometry for the galaxy cluster components in the system.

\begin{figure}[ht!]
    \centering
    \includegraphics[width=0.45\textwidth]{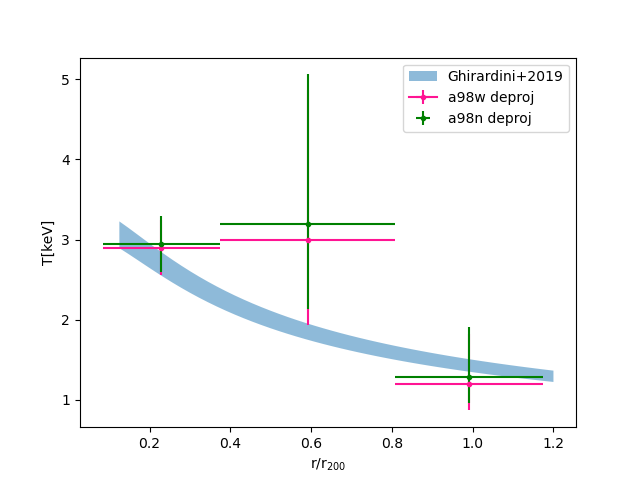}
    \caption{The deprojected temperature profile of A98N for the sectors to the north and west shown in Figure~\ref{fig:mosaic}. The pink points are for the sectors to the west of A98N and the green points are for the sectors to the north of A98N. The blue line is the ``Universal" temperature profile presented in \cite{ghirardini2019}}
    \label{fig:a98NuniversalTempProf}
\end{figure}

The {\it Suzaku} temperature profile of A98N in partial \deleted{aziumth} \edit1{azimuth} (see sector regions in e.g. Figure~\ref{fig:mosaic} (Left)) is shown in Figure~\ref{fig:a98NuniversalTempProf} as well as the ``Universal" temperature profile \citep[][]{ghirardini2019} expected based upon the average electron temperature of the system. While not all of the error constraints are tight, the temperature profile of A98N is not only consistent in partial azimuth, but also seems to adhere to expectations of the temperature profile based on average cluster temperature (cluster $T_{500}$ measurements with {\it Suzaku} are presented in Table~\ref{tab:objects}).

\begin{figure}[ht!]
    \centering
    \includegraphics[width=0.45\textwidth]{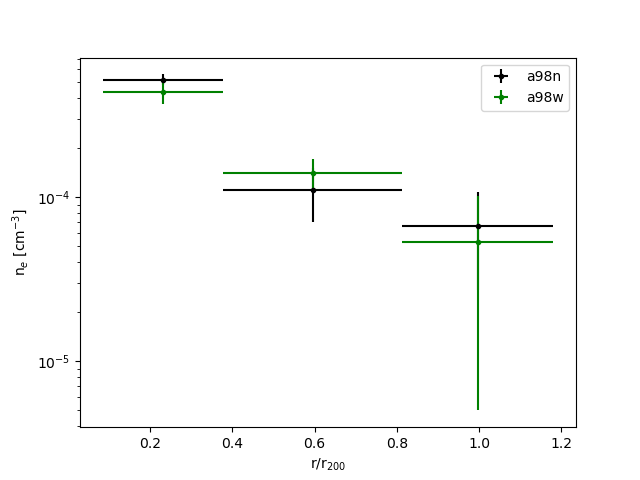}
    \caption{The electron density profile of the annuli to the north and west of A98N (see e.g. Figure~\ref{fig:mosaic} (Left))}.
    \label{fig:edensityprofNW}
\end{figure}

The electron density profile of A98N in the northern and western sectors is shown in Figure~\ref{fig:edensityprofNW}. The profiles are consistent with each other within errors.

\begin{figure}[ht!]
    \centering
    \includegraphics[width=0.45\textwidth]{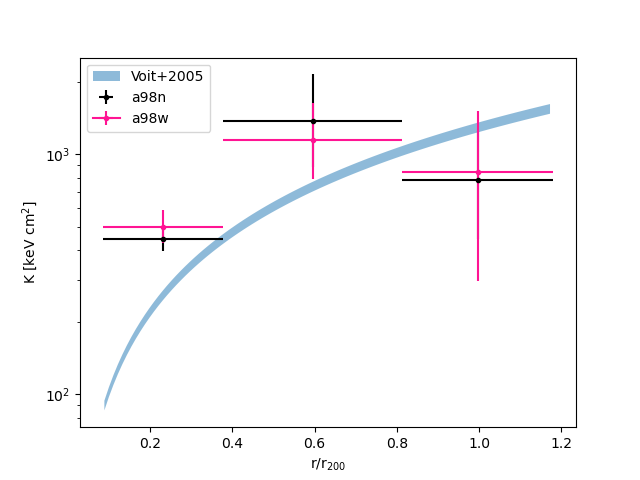}
    \caption{The blue line is the entropy profile expected from pure gravitational collapse \citep{voit} for A98N. The black points are for the sector regions to the north of A98N and the pink points are for the sector regions to the west of A98N shown in Figure~\ref{fig:mosaic}.}
    \label{fig:a98nent}
\end{figure}

The deprojected entropy profiles for A98N in the northern and western sectors, for the regions shown in e.g. Figure~\ref{fig:mosaic} (Left), are shown in Figure~\ref{fig:a98nent}. Here, entropy is defined as $K = kT_e n_e^{-2/3}$, where $k$ is Boltzmann's constant, $T_e$ is the measured electron temperature, and $n_e$ is the derived electron density assuming a region appropriate volume as described in Equation~\ref{eq:density}. \edit1{The statistical uncertainty in the normalization for the self-similar entropy profile in Figure~\ref{fig:a98nent}, and the normalization of the temperature profile presented in Figure~\ref{fig:a98NuniversalTempProf} are included as the shaded blue region. Unknown biases such as instrumental calibrations \citep[i.e.][]{schellenberger2015} could affect any conclusions derived from these figures.}

We find that, in both sectors, the entropy values for A98N are consistent with the self-similar prediction of pure gravitational collapse, \edit1{based upon spherical hydrodynamic (SPH) simulations of galaxy clusters}, near the virial radius, in which entropy $K\propto r^{1.1}$ \citep{voit}. There is some hint of entropy flattening, however the errors on the entropy are too large to say definitively. This is in agreement with similar studies of lower mass systems \citep[e.g.][]{su,bulbul,sarkar} even though A98N is most likely undergoing a merging event \citep[][]{paterno-mahler,sarkarA98}. At small cluster radii ($r < r_{500}$), the entropy for A98N is above that expected for self-similarity. This excess is consistent with what is seen in other systems, particularly lower mass systems \citep[][]{sun}, and is likely due to the effect of non-gravitational processes in the central region (e.g. AGN feedback \citep[e.g.][]{osullivan}).

\subsection{Large-Scale Filament} \label{subsec:filres}

The box region to the north of A98N shown in Figure~\ref{fig:mosaic} \edit1{Left} is used to investigate larger-scale structure beyond $r_{200}$ of A98N. The excluded region within the box does not appear to be a point source, and also does not appear to be associated with the A98 system. Inspecting images from the DSS and SDSS reveals no obvious clustering of galaxies. 
There are not enough photons available to reliably model the spectrum of this source. This source could be a background galaxy cluster or faint group, although it would require further observations to determine its nature. \edit1{A 1D-$\beta$ model is fit to the circular region encompassing the possible background cluster, with best fit parameters $\beta \approx 1.3$ and $r_c \approx 2'$. This model is extrapolated to compare the expected surface brightness, to the measured surface brightness in the rest of the box region shown in Figure~\ref{fig:mosaic} (left panel). The measured surface brightness is larger than the expected surface brightness by approximately a factor of 4.}

This region was fitted with an absorbed APEC model. The best fit values for this box region, excluding the possible background cluster, yield a \edit1{projected} temperature of \deleted{$kT = 0.58_{-0.2}^{+1.3}$~keV, electron density $n_e = {8.3 \times 10^{-5}}^{+1.3 \times 10^{-4}}_{-7.4 \times 10^{-5}}$~cm$^{-3}$, and a projected entropy K = $326_{-267}^{+828}$~keV cm$^2$} \edit1{$kT = 0.11_{-0.02}^{+0.01}$~keV, projected electron density $n_e = {7.6 \times 10^{-5}}^{+3.6 \times 10^{-5}}_{-3.6 \times 10^{-5}}$~cm$^{-3}$, and a projected entropy K = $61_{-22}^{+20}$~keV cm$^2$} assuming a cylindrical volume with length \deleted{$l=1.2$} \edit1{$l = 0.62$}~Mpc and radius of \deleted{$r=0.5$} \edit1{$r = 0.61$}~Mpc, and filling factor of 1 for the density measurement. These values are consistent with the dense end of the WHIM, \edit1{where the ICM in the outskirts of the cluster interact with the large-scale structure filament \citep[e.g.][]{dolag,werner}}, \edit1{especially if the system is not oriented in the plane of the sky.}
 
We compare a similar region to the west, and fit the region with the best fit parameters for the model of the northern region. The fit for the northern box region yields a $\chi^2$~of \deleted{$353.4$ with $318$} \edit1{$289.24$~with $272$} d.o.f. We find that the western comparison region is consistent with background only with a fit that gives a $\chi^2$~of \deleted{$540.23$ with $320$} \edit1{$555.21$ with $321$} d.o.f when the data is fit to the best fit model of the northern region. \edit1{Additionally, this region is fit with a background only model for comparison. The model with the extra APEC component ($\chi^2$/d.o.f = $289.25/273$) is marginally preferred ($p = 0.06$) over the background only model ($\chi^2$/d.o.f = $293/274$).} Such detections of this material are very rare \citep[e.g.][]{bulbul,werner}. This detection is consistent with the detection of larger-scale structure in the Abell 1750 system, another low mass triple cluster system \citep[][]{bulbul} similar to A98. With the advent of {\it eRosita} and future X-ray missions (e.g. {\it Athena, Lynx}), such detections should become more common.

\subsubsection{A98N-A98S Bridge} \label{subsec:a98na98sBridge}

To investigate whether the apparent surface brightness enhancement in between A98N and A98S is due to two cluster halos overlapping, we compare the combined surface brightness profiles of A98N and A98S to the emission across the bridge, as done in \citep[][]{paterno-mahler,sarkarA98} .

\begin{figure}[ht!]
    \centering
    \includegraphics[width=0.45\textwidth]{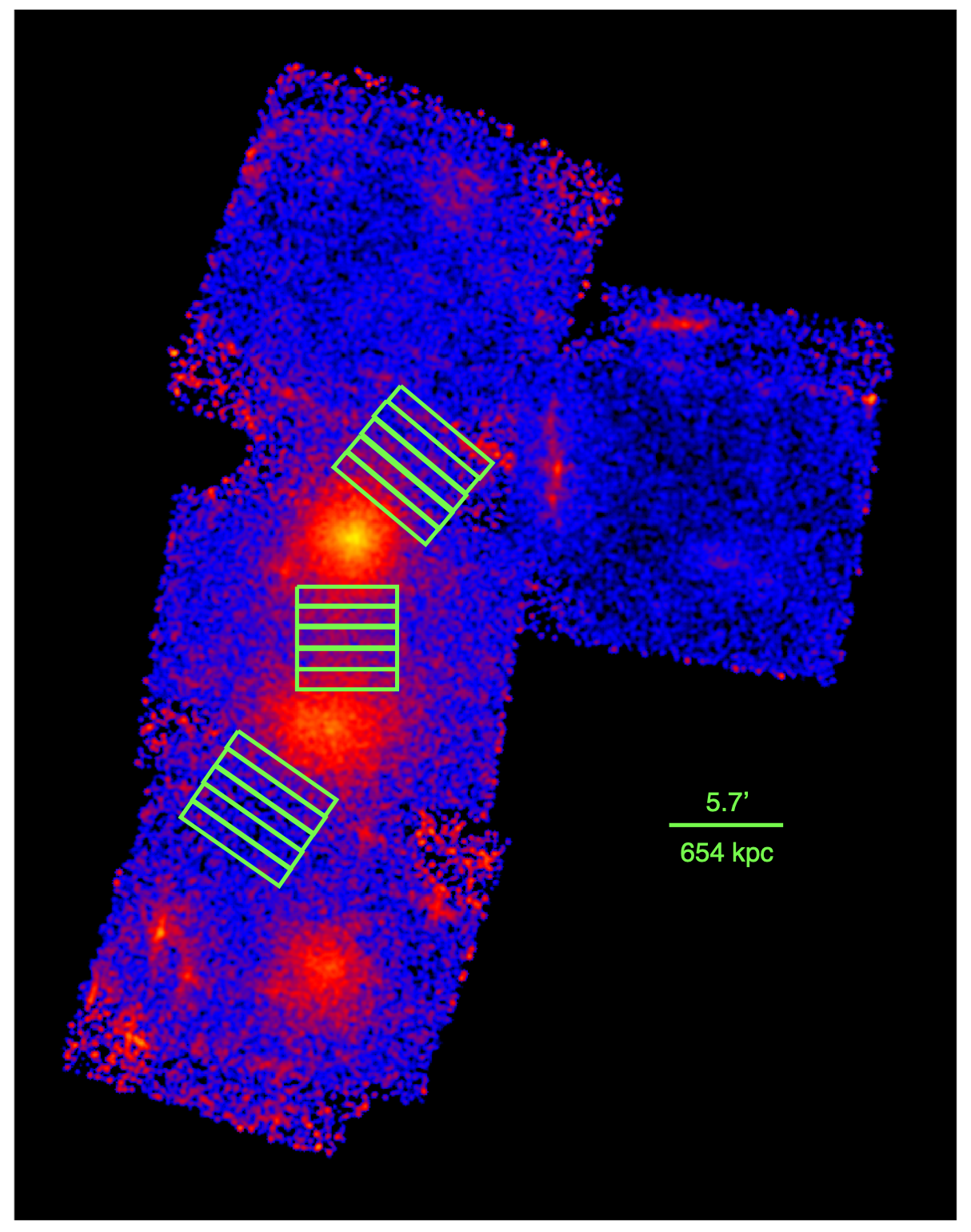}
    \caption{Same as Figure~\ref{fig:mosaic}: Left, but instead smoothed with a $12.5\arcsec$~Gaussian. The rectangular annuli to the north of A98N and to the south of A98S are used to compare the combined surface brightness of the two clusters to the box regions in the bridge region (see Figure~\ref{fig:a98na98sFilComp}).}
    \label{fig:a98na98sBridgeProfile}
\end{figure}

We combine the background subtracted surface brightness profiles of the box annuli to the north of A98N and to the south of A98S shown in Figure~\ref{fig:a98na98sBridgeProfile}. These box annuli are oriented to avoid the large-scale and intercluster filaments in the system. We then compare these values to the surface brightness in the apparent bridge regions connecting A98N and A98S shown in Figure~\ref{fig:a98na98sBridgeProfile}. 

\begin{figure}[ht!]
    \centering
    \includegraphics[width=0.45\textwidth]{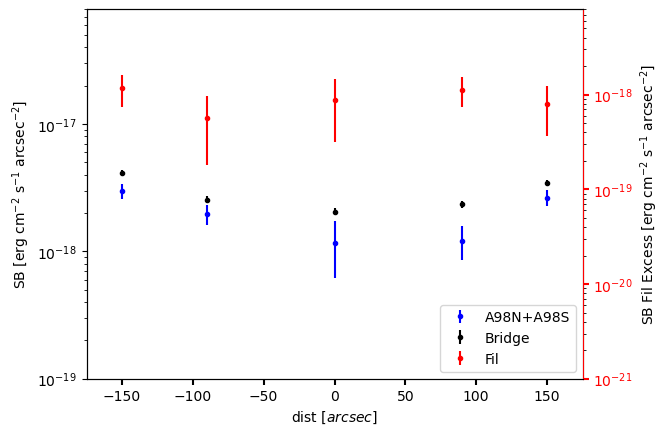}
    \caption{The surface brightness profile with XIS3 in the $0.5-2.0$~keV range, and $1\sigma$~errors for the filamentary region shown in Figure~\ref{fig:a98na98sBridgeProfile} is shown in black points, using the y-axis scale on the left. The corresponding combined rectangular annuli shown in Figure~\ref{fig:a98na98sBridgeProfile} for A98N and A98S are shown for comparison with the blue points. The filament excesses, are shown in red, using the y-axis scale on the right. The zero point on the x-axis is the midpoint of the tentative bridge connecting A98N to A98S.}
    \label{fig:a98na98sFilComp}
\end{figure}

The combined ICM surface brightness profile as compared to the bridge, and the resulting filament excess is presented in Figure~\ref{fig:a98na98sFilComp} where the zero point on the x-axis is the midpoint of the region connecting A98N and A98S. The emission across the A98N-A98S bridge appears to be slightly enhanced, with a marginal ($2.2\sigma$) detection of excess filament emission. This excess emission is also seen at a higher significance with combined {\it Suzaku} and deeper {\it Chandra} observations \citep[][]{sarkarA98} than are presented in this paper.

\subsubsection{Bridge Thermodynamic Properties}\label{subsec:bridgeProp}

\begin{figure}[ht!]
    \centering
    \includegraphics[width=0.45\textwidth]{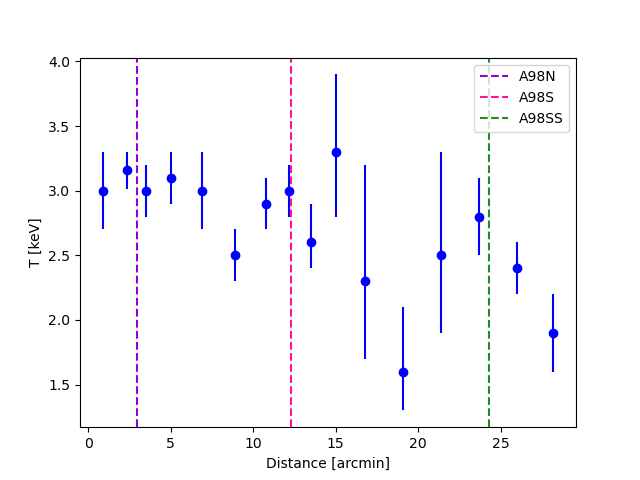}
    \caption{The temperature profile for the box regions across the A98 system shown in Figure~\ref{fig:mosaic}. The profile starts from the north on the left of the plot. The purple, pink, and green dashed lines mark the center of A98N, A98S, and A98SS respectively.}
    \label{fig:a98bridgetemp}
\end{figure}

We measure the temperature, electron density, entropy, and metallicity across the A98 system with the box regions shown in e.g. Figure~\ref{fig:mosaic} (Right). The temperature profile from north to south is shown in Figure~\ref{fig:a98bridgetemp}, and is relatively flat across A98N and A98S before dipping in between A98S and A98SS. When compared to {\it Chandra}, the temperature profiles are consistent.

We find that in the intercluster region between A98N and A98S that a two temperature model is preferred (Table~\ref{tab:2temp}). We freeze the metallicity at its best fit value found from the one temperature fit, and freeze the metallicity of the second temperature component to $\deleted{z}\edit1{Z} = 0.2$~Z$_\odot$.

\begin{table}[]
\label{tab:2temp}
\centering
\caption{Parameters for a 1 temperature and 2 temperature model of the intercluster region between A98N and A98S.}
\begin{tabular}{@{}ccccc@{}}
\toprule
Model  & $T_1$ {[}keV{]}     & $T_2$ {[}keV{]}    & $\chi^2$ & d.o.f \\ \midrule
1T APEC & $2.7_{-0.2}^{+0.3}$ & -                   & 267.3   & 225 \\
2T APEC & $4.0_{-1.0}^{+6.5}$ & $1.3_{-0.3}^{+0.3}$ & 249.9    & 224 \\ \bottomrule
\end{tabular}
\end{table}

\begin{figure}[ht!]
    \centering
    \includegraphics[width=0.45\textwidth]{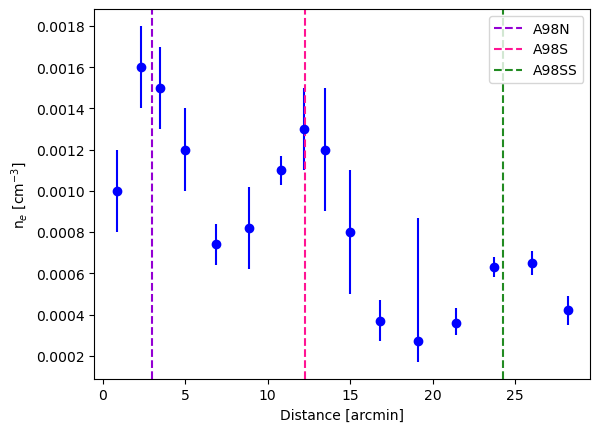}
    \caption{Inferred electron density profile across the A98 system. The profile is starting from the north on the left of the plot. The purple, pink, and green dashed lines mark the center of A98N, A98S, and A98SS respectively.}
    \label{fig:nevxallfil}
\end{figure}

The inferred electron density (Equation~\ref{eq:density}) for the regions across the bridge is presented in Figure~\ref{fig:nevxallfil}. A cylindrical geometry is assumed for the regions in this profile. Assuming a box geometry for the regions across the bridge yields the same result within errors; therefore neither geometry is preferred. There is a steady decline in the electron density profile with local peaks corresponding to the cluster centers.

\begin{figure}[ht!]
    \centering
    \includegraphics[width=0.45\textwidth]{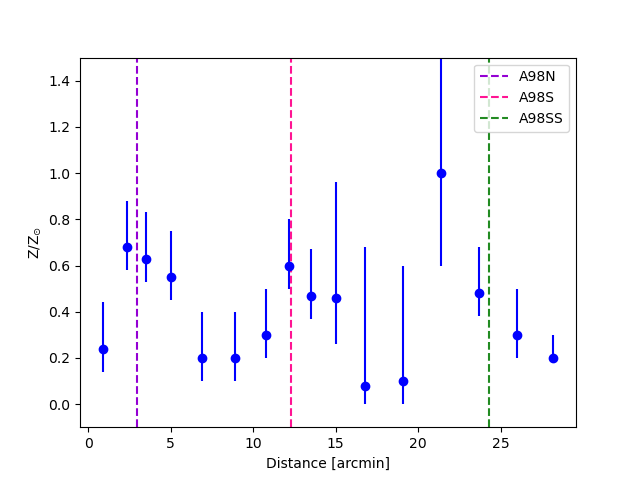}
    \caption{The metallicity profile across the A98 bridge. The profile is starting from the north on the left of the plot. The purple, pink, and green dashed lines mark the center of A98N, A98S, and A98SS respectively.}
    \label{fig:metalprof}
\end{figure}

The metallicity profile across the bridge is shown in Figure~\ref{fig:metalprof}. This profile suggests that the metallicity of the ICM is radius dependent, with a higher metallicity towards the center of the clusters, and decreases with radius.

\subsubsection{Filament Orientation} \label{subsubsec:filOrient}

\begin{figure}[ht!]
    \centering
    \includegraphics[width=0.45\textwidth]{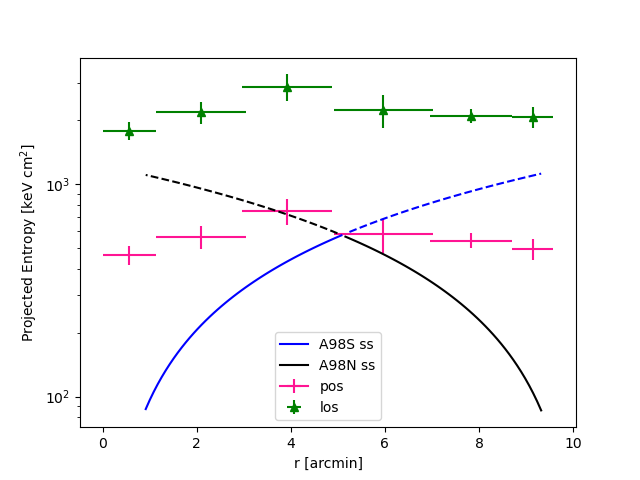}
    \caption{The entropy profile from north to south across the bridge between A98N and A98S. The blue line is the expected universal entropy profile for A98N. The black line is the expected universal entropy profile for A98S. The pink points are the inferred entropy for a filament orientation in the pos. The green triangles are for a filament orientation along the los. A filament orientation closer to the pos is preferred, as the los entropy values are $\sim2$ times higher than what is expected in the outskirts of both subclusters.} 
    \label{fig:a98bridgeent}
\end{figure}

The cyan box regions shown in e.g. Figure~\ref{fig:mosaic} (Right) between A98N and A98S can be used to explore the entropy, presumed geometry, and inclination of the system. The entropy profile for the box regions between A98N and A98S is shown in Figure~\ref{fig:a98bridgeent}. We only include the regions from the center of A98N to the center of A98S in this entropy profile to investigate the filament orientation (cyan regions in e.g. Figure~\ref{fig:mosaic} (Right)).

The two extremes for filament orientation: along the line of sight (los) and in the plane of the sky (pos) are investigated (see Figure~\ref{fig:a98bridgeent}).  At the midpoint of the filament, the measured entropy is already approximately the expected ICM value at this radius if the filament is in the pos. This means that either the filament is in the pos, or the entropy of the gas has somehow been increased (e.g., due to a merger shock \citep[][]{sarkarA98}). The filament is more likely inclined close to the pos, as a los filament orientation yields entropy values well above what is expected from the self-similar entropy profiles of the clusters \citep{voit}. Furthermore, a line of sight orientation of the system yields entropy values $\sim2$ times higher than what is expected in the outskirts of both subclusters. These measurements rule out a significant contribution from WHIM emission in this region. The reported entropy values similarly assume a cylindrical geometry for the bridge regions, and assuming a box geometry yields a similar result within errors.

\section{Conclusions and Summary} \label{sec:conclusionsSumm}

In this paper we present results from an analysis of {\it Suzaku, Chandra} and XMM-{\it Newton} observations of the diffuse emission in the A98 system. 

We find the following:

\begin{itemize}
    \item The entropy profiles in northern and western sectors, along and away from the merger axis and the putative large scale structure filament, for A98N generally agree with each other, and with the self-similar expectation in the virialization region. This is consistent with previous suggestions \citep[e.g.][]{su,bulbul} that lower mass clusters and groups adhere more closely to self-similar expectations in their outskirts, in contrast with what is seen in most massive systems.
    \item The region to the north of A98N, \deleted{at and} beyond $r_{200}$, was found to have a temperature, density, and entropy consistent with those expected for the dense end of the WHIM. We find a temperature of \deleted{$kT =0.58_{-0.2}^{+1.3}$~keV, electron density $n_e = {8.3 \times 10^{-5}}^{+1.3 \times 10^{-4}}_{-7.4 \times 10^{-5}}$~cm$^{-3}$, and entropy $K=326^{+828}_{-267}$\edit1{keV cm$^2$}} \edit1{$kT = 0.11_{-0.02}^{+0.01}$~keV, projected electron density $n_e = {7.6 \times 10^{-5}}^{+3.6 \times 10^{-5}}_{-3.6 \times 10^{-5}}$~cm$^{-3}$, and a projected entropy K = $61_{-22}^{+20}$~keV cm$^2$} for this region. \deleted{However, the errors on these values are large.} The presence of similar emission at the same radius to the west, away from the putative large scale structure filament, is ruled out.  This serves as further evidence that the system is consistent with the expectation that the merger axis lies along a large-scale structure filament. A similar result was found in the colinear triple system Abell~1750 \citep[][]{bulbul}.  These measurements provide tantalizing evidence for the presence of a larger-scale structure, with the diffuse WHIM connecting to the cluster outskirts along cosmic filaments.
    \item When comparing the surface brightness of the A98N-
    A98S bridge regions to the combined surface brightness profiles of the two overlapping halos, a nominal $2.2\sigma$ excess in bridge emission is detected. This detection is suggestive of the presence of an intercluster filament in-between the two clusters. Additionally, there is evidence of two-phase plasma in this region; the lower temperature component ($kT \sim 1$~keV) is consistent with the dense end of the WHIM.
    \item Comparing the entropy profile of the A98N-A98S bridge to that of the self-similar expectations for A98N and A98S reveals that the system is likely inclined closer to the pos. This suggests that the clusters are interacting with each other as they are well within each other's virial radius in projection.
\end{itemize}

In this study, a picture similar to the large-scale structure seen in cosmological simulations starts to emerge. {\it Suzaku} is a powerful tool for studying diffuse ICM emission at large cluster radii, but is no longer functional and available for future observations. The next generation of X-ray telescopes such as {\it eRosita, XRISM, Athena}, and {\it Lynx}  will provide a wealth of information on the diffuse ICM in the virialization region of galaxy clusters and the surrounding large-scale structure \citep[e.g.][]{reiprich}. The ability to study increasingly low surface brightness cluster outskirt emission will help to answer key questions about the physical processes occurring in these interface regions. This will lead to a new era of synergy between observation and simulation in the pursuit to understand cosmology and the physics that govern the observable Universe.

\acknowledgments

{Support for GEA was partially provided by Chandra X-ray Observatory grants GO9-20118X, GO0-21113X, 80NSSC18K0915, 80NSSC19K0865, and GO9-20112X and by the STScI grant HST-GO-15610.001. Support for SWR provided by the Chandra X-ray Center through NASA contract NAS8-03060, the Smithsonian Institution, and by the Chandra X-ray Observatory grant GO3-14134X.}

\bibliographystyle{aasjournal}
\bibliography{A98bib}

\end{document}